\documentclass[lettersize,journal]{IEEEtran}
\usepackage{amsmath,amsfonts}
\usepackage[noend]{algpseudocode} % No need for end statements
\usepackage{algorithm} % For using the algorithm environment
\usepackage{amsmath}
\usepackage{array}
\usepackage[caption=false,font=normalsize,labelfont=sf,textfont=sf]{subfig}
\usepackage{textcomp}
\usepackage{stfloats}
\usepackage{url}
\usepackage{verbatim}
\usepackage{graphicx}
\usepackage{subcaption}

\def\BibTeX{{\rm B\kern-.05em{\sc i\kern-.025em b}\kern-.08em
    T\kern-.1667em\lower.7ex\hbox{E}\kern-.125emX}}
\usepackage{balance}
\begin{document}
\title{Converter-Grid Interaction Stability Guaranteed Safe Deep Reinforcement Learning for Energy Storage Systems in Grid Frequency Support}
\author{Fei Liu, \textit{Graduate Student Member, IEEE}, Mengfan Zhang, \textit{Member, IEEE}, Zhipeng Li, \textit{Graduate Student Member, IEEE}, Frede Blaabjerg, \textit{Fellow, IEEE}, Qianwen Xu, \textit{Senior Member, IEEE}
\thanks{This work was supported by Swedish Research Council Starting Grant under Project 2021-04434 and by Swedish Energy Agency under Project P2022-00775.

This work has been submitted to the IEEE for possible publication. Copyright may be transferred without notice, after which this version may no longer be accessible.
}}

\markboth{Journal of \LaTeX\ Class Files,~Vol.~, No.~, Month~Year}%
{How to Use the IEEEtran \LaTeX \ Templates}

\maketitle

\begin{abstract}
The growing integration of converter interfaced renewable energy resources (RESs) intensifies stability challenges. Energy storage system (ESS) can provide fast and flexible frequency support to mitigate frequency deviations. However, the interface converter of ESS may encounter converter-grid interaction stability issues.
This paper proposes a converter-grid interaction stability guaranteed safe DRL (CIS-DRL) method for ESS integrated power systems to achieve frequency regulation.
We first obtain a double DNN-based stability region to identify the guaranteed converter-grid interaction stability. Next, a novel converter-grid interaction stability Safe-TD3 (CIS-STD3) algorithm is designed that integrates a stability feasibility projection layer to map unsafe actions into stable action set before execution, enforcing converter-grid interaction stability as a hard constraint throughout learning process. 
The proposed approach enables ESS for grid frequency support with 100\% converter-grid interaction stability without violations. 
Experimental results show that the proposed CIS-DRL method achieves improved frequency regulation performance while preventing unstable operating points, demonstrating its practical applicability for real time ESS frequency support.

\end{abstract}


\section{Introduction}

\IEEEPARstart{T}{he} stability of power systems has traditionally been analyzed the electromechanical dynamics of synchronous generators, addressing relatively slow phenomena such as rotor angle, voltage, and frequency stability \cite{b1}. However, with the growing integration of renewable energy resources (RESs) and energy storage systems (ESSs), power systems are increasingly operated with converter interfaced generators (CIGs) and reduced synchronous inertia. The increasing reliance on converter controls influences the system dynamic behavior and brings new converter driven stability concerns \cite{b2}. The fast control dynamics of CIGs, such as the phase-locked loop (PLL) and the
inner-current control loops can result in cross couplings with both the grid impedance and the network dynamics, causing converter–grid interaction stability issues over wide frequency ranges \cite{b3}. 

ESSs are widely used for fast frequency support in low inertia systems, and many control methods have been proposed to reduce frequency deviations.
In \cite{b8}, a frequency constrained unit commitment framework is developed by coordinating wind turbines and ESS to enhance frequency security. A hierarchical coordinated control strategy is proposed in \cite{b9} to improve ESS performance in secondary frequency regulation with coordinated power allocation and state of charge (SOC) management. A virtual inertia emulator integrating model predictive control is designed in \cite{b10} to improve grid frequency regulation.
However, these studies only focus on frequency performance and operational boundaries, while converter–grid interaction stability is rarely modeled and enforced during operation.

To characterize converter–grid interaction stability, eigenvalue-based small signal analysis and impedance-based stability theory have been widely used. These methods can reveal interaction modes and quantify stability margins via impedance ratios or Nyquist-type criteria \cite{r1}. Recent measurement-oriented modeling and online impedance identification further enable real-time margin tracking using grid and converter evaluate impedance estimation \cite{r2,r3}. However, these methods are mainly used for offline stability analysis and evaluation, rather than being integrated into real-time operation level decision making.

To consider stability problems in operation, several stability-constrained optimization methods have been proposed \cite{b11}-\cite{b14}. A two stage robust strategy jointly schedules flexible resources and tunes droop gains to guarantee small signal stability in inverter dominant microgrids under renewable and load uncertainties \cite{b12}. Other studies embed small signal stability constraints into optimal power flow by utilizing Lyapunov conditions or eigenvalue based indices, and solve the resulting problem through iterative verification or relaxation procedures \cite{b13,b14}. However, converter-grid interaction stability is absent from these operation level scenarios. The constraints are derived from generic eigenvalue or Lyapunov indices and therefore do not capture impedance coupling interaction boundaries during real-time operation. Moreover, existing stability-constrained optimization typically relies on explicit dynamic models and iterative stability checks, leading to computationally heavy and nonconvex formulations that are difficult to deploy when detailed converter and inner-loop information are unavailable \cite{a3}.

Data-driven and artificial intelligence (AI) approaches have been explored to estimate complex stability boundaries and reduce reliance on detailed dynamic models \cite{b15}. Existing learning based studies cover two main directions. One direction uses deep reinforcement learning (DRL) to assist converter control and gain tuning under nonlinear and uncertain conditions \cite{b16,b17}, which primarily improves local dynamic performance. The other direction learns stability related constraints and embeds them into scheduling or dispatch models \cite{b18,b20}. However, these approaches are commonly trained offline for fixed system configurations and largely target conventional stability constraints, not consider converter inner-loop dynamics, and cannot address converter–grid interaction stability under wide operating variations.

Under wide operating variations, real-time operation requires DRL policies with hard constraints, covering many operational limits such as converter current and voltage limits, ESS power outputs and SOC bounds. Recent safe DRL studies have introduced safety filters and action projection mechanisms to correct infeasible actions before execution \cite{b22,b23}, and have demonstrated near zero violations in applications such as Volt/Var control of active distribution systems \cite{a5}. Safe DRL has also been investigated for grid forming inverter frequency regulation by constructing Lyapunov based safe regions of states and actions \cite{a6}. However, the enforced hard constraints in these works remain conventional and only focus on voltage magnitude limits or stability conditions tied to a specific inverter control loop, and converter-grid interaction stable operation is therefore not guaranteed at the system operation level.

To fill the research gaps, this article proposes a converter-grid interaction stability guaranteed safe DRL (CIS-DRL) method for ESS integrated power systems. The proposed method first obtains the double DNN-based stability region of the converter-grid interaction system to identify the stable operation set for ESS frequency support, and support real-time stability monitoring without detailed converter inner-loop information. Next, a CIS-DRL algorithm is designed by integrating a CIS projection layer into Safe-Twin Delayed Deep Deterministic Policy Gradient (TD3), which maps any unstable action into the stable action set before being applied to the system. The proposed approach enables optimal frequency regulation with guaranteed CIS under wide variations of operating points.

The remainder of this paper is organized as follows:
Section II introduces the problem formulation for ESS frequency regulation. Section III presents the stability-constrained frequency regulation model, CIS-DRL based control environment, Safe-TD3 algorithm with the CIS projection layer and its implementation. Section IV provides the case study and the experimental test results to validate the effectiveness of the proposed method. Finally, Section V concludes this paper.

\section{Problem Formulation}

\begin{figure}[htbp]
\centerline{\includegraphics[width=0.35\textwidth]{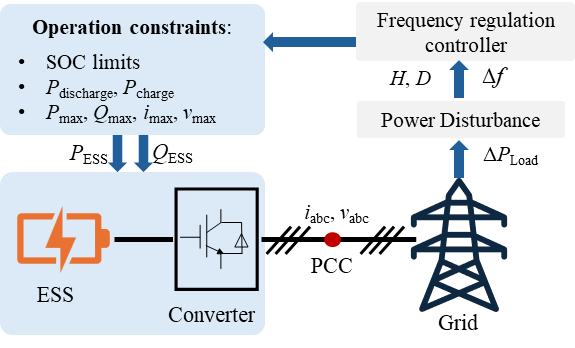}}
\caption{Frequency regulation by ESS.}
\label{frequencyregulationESS}
\end{figure}

This section formulates the operation level frequency regulation problem of a grid-connected ESS in low-inertia grids. 
The regulation objective over the control horizon $\mathcal{T}$ is defined as:
\begin{equation}
\min \ \sum_{t\in\mathcal{T}} \big|\Delta f_t\big|,
\label{eq:obj_freq}
\end{equation}
where $\Delta f_t$ is the system frequency deviation. The frequency dynamics follow the discretized swing equation with damping:
\begin{equation}
\Delta f_{t+1}
=\Delta f_t
+\frac{\Delta t}{2H}\Big(\Delta P_t - D\,\Delta f_t - P_{\mathrm{ESS}}(t)\Big),
\label{eq:swing_discrete}
\end{equation}
where $\Delta t$ is the sampling period, $H$ is the inertia constant, $D$ is the damping coefficient, and $\Delta P_t$ denotes the net power imbalance excluding the ESS contribution.

In addition to the frequency regulation objective, the problem formulation is subject to the ESS converter operational boundary specifications, including charging and discharging power limits, SOC bounds, converter current constraints, and voltage operating limits.
The charging and discharging constraints are:
\begin{equation}
P_{\mathrm{ESS}}(t)=P_{\mathrm{dis}}(t)-P_{\mathrm{ch}}(t),
\label{eq:p_split}
\end{equation}
where $P_{\mathrm{dis}}(t)$ and $P_{\mathrm{ch}}(t)$ denote discharging and charging powers:
\begin{align}
0 \le P_{\mathrm{dis}}(t) &\le u(t)\,P_{\max},
\label{eq:pdis_lim}\\
0 \le P_{\mathrm{ch}}(t) &\le \big(1-u(t)\big)\,P_{\max},
\label{eq:pch_lim}\\
u(t) &\in \{0,1\},
\label{eq:u_binary}
\end{align}
where $P_{\max}$ is the rated power and $u(t)$ indicates the discharging mode. The SOC dynamics satisfy:
\begin{equation}
\mathrm{SOC}(t{+}1)=\mathrm{SOC}(t)
+\frac{\eta_c P_{\mathrm{ch}}(t)-\frac{1}{\eta_d}P_{\mathrm{dis}}(t)}{E}\Delta t,
\label{eq:soc_dyn}
\end{equation}
where $\eta_c$ and $\eta_d$ are charging and discharging efficiency, and $E$ is energy capacity. The SOC is bounded by:
\begin{equation}
\mathrm{SOC}_{\min}\le \mathrm{SOC}(t)\le \mathrm{SOC}_{\max}.
\label{eq:soc_bounds}
\end{equation}

The converter current magnitude limit is enforced as:
\begin{equation}
\sqrt{\left(\frac{P_{\mathrm{ESS}}(t)}{v_d(t)}\right)^2
+\left(\frac{Q_{\mathrm{ESS}}(t)}{v_d(t)}\right)^2}\le i_{\max},
\label{eq:current_limit}
\end{equation}
where $i_{\max}$ is the current rating. 

The voltage operating range by grid standards:
\begin{equation}
v_{\min}\le v_d(t)\le v_{\max}.
\label{eq:voltage_limit}
\end{equation}

Let $\mathcal{A}_{\mathrm{op}}(t)$ with \eqref{eq:p_split}--\eqref{eq:voltage_limit} denote the operation limits that ESS power commands action set $\big(P_{\mathrm{ESS}}(t),Q_{\mathrm{ESS}}(t)\big)$ should satisfy. 
However, $\mathcal{A}_{\mathrm{op}}(t)$ does not include converter-grid interaction stability, and frequency support actions may drive the converter operating point into unstable regions due to the interactions of ESS converter and grid. 
Since the associated stability boundary is highly nonlinear and nonconvex, conventional optimization with iterative feasibility checks becomes difficult to deploy online. Simply relaxing the stability constraint into a soft penalty in the objective may also result in unsafe intermediate solutions and can not guarantee the converter-grid interaction stability. To overcome these limitations, the following section adopts the proposed CIS-DRL to enforce interaction stability
during real-time frequency regulation.

\section{CIS-DRL based Frequency Regulation with Converter-Grid Interaction Stability Region Integrated}

In this section, we propose CIS-DRL to achieve ESS for frequency support with guaranteed converter-grid interaction stability. 
The section first introduces the stability region evaluation, which determine whether a candidate ESS operating point is stable. 
Then, the CIS-DRL environment is defined to model the multi-step frequency regulation process. 
Based on the stability region and the environment, a converter-grid interaction stability Safe-TD3 (CIS-STD3) algorithm is developed. The converter-grid interaction stability projection layer is integrated into Safe-TD3, in which any stability violating action is projected into the stable action set before execution. 
Finally, the section shows the overall method.

\subsection{Converter-grid interaction stability region generation}

The converter-grid interaction stability is operating point dependent, and safe frequency regulation requires fast characterization of the corresponding stability region under wide operating variations \cite{b24}. 
To describe the converter operating condition relevant to stability assessment, the operating point at the point of common coupling (PCC) is represented in the synchronous $dq$ frame. The three-phase variables are transformed into approximately constant steady-state quantities, which enables a compact operating point description and tractable stability analysis:
\begin{equation}
\boldsymbol{o}(t)=\big[v_d(t),\,v_q(t),\,i_d(t),\,i_q(t)\big]^{\mathrm T},
\label{eq:op_full}
\end{equation}
where $v_d(t)$ and $v_q(t)$ are the $d$- and $q$-axis components of the PCC voltage, and $i_d(t)$ and $i_q(t)$ are the corresponding converter output currents. With a PLL aligning the $d$-axis to the PCC voltage, $v_q(t)$ is regulated approaches zero.

Let $P_{\mathrm{ESS}}(t)$ and $Q_{\mathrm{ESS}}(t)$ denote the instantaneous active and reactive powers delivered by the ESS converter at time $t$:
\begin{align}
P_{\mathrm{ESS}}(t) &= i_d(t)\,v_d(t) + i_q(t)\,v_q(t),
\label{eq:pq_P}\\
Q_{\mathrm{ESS}}(t) &= i_d(t)\,v_q(t) + i_q(t)\,v_d(t),
\label{eq:pq_Q}
\end{align}
and under $v_q(t)=0$, the decoupled relations are obtained as:
\begin{align}
P_{\mathrm{ESS}}(t) &= v_d(t)\,i_d(t),
\label{eq:p_dec}\\
Q_{\mathrm{ESS}}(t) &= v_d(t)\,i_q(t),
\label{eq:q_dec}
\end{align}

Converter-grid interaction stability is quantified by a stability margin:
\begin{equation}
m\big(\boldsymbol{o}(t)\big):\mathbb{R}^3 \rightarrow \mathbb{R},
\label{eq:margin_map}
\end{equation}
where $m\big(\boldsymbol{o}(t)\big)$ represents the stability margin at operating point $\boldsymbol{o}(t)$. The stability boundary is characterized by $m(\boldsymbol{o})=0$, and a larger positive margin indicates a larger distance to the boundary. The stability admissible region in the operating point space is defined as:
\begin{equation}
\mathcal{O}_{\mathrm{stab}}=\big\{\boldsymbol{o}:\ m(\boldsymbol{o})\ge \tau\big\},
\label{eq:O_stab}
\end{equation}
where $\tau \ge 0$ is a prescribed safety margin.

Since the control decision is made in the power command space, the admissible region in \eqref{eq:O_stab} induces a stability action set at time $t$ through \eqref{eq:p_dec}--\eqref{eq:q_dec}:
\begin{equation}
\mathcal{A}_{\mathrm{stab}}(t)=\Big\{\ 
m\!\Big(\big[v_d(t),\,\frac{P_{\mathrm{ESS}}(t)}{v_d(t)},\,\frac{Q_{\mathrm{ESS}}(t)}{v_d(t)}\big]^{\mathrm T}\Big)\ge \tau\Big\},
\label{eq:A_stab}
\end{equation}
The set $\mathcal{A}_{\mathrm{stab}}(t)$ is operating point dependent and generally nonlinear and nonconvex, because it combines the complex converter-grid interaction dynamics embedded in the stability margin and the time varying PCC voltage $v_d(t)$.

At the operation level, an explicit analytical expression of $m(\cdot)$ is typically unavailable due to limited access to converter internal control details. Therefore, $m(\cdot)$ is treated as a black-box stability margin evaluation map constructed from measurement based stability assessment over multiple operating points.
Combined with the operational feasibility set $\mathcal{A}_{\mathrm{op}}(t)$ defined in Section~II, the admissible decision set is:
\begin{equation}
\mathcal{A}(t)=\mathcal{A}_{\mathrm{stab}}(t)\cap \mathcal{A}_{\mathrm{op}}(t).
\label{eq:A_total}
\end{equation}
which will be utilized in the proposed CIS-DRL method.

\subsection{Environment formulation for converter-grid interaction stability guaranteed frequency regulation}

The frequency regulation problem with converter-grid interaction stability feasibility is formulated as a constrained Markov decision process. 
At each time step $t$, the controller observes the system state $s_t\in\mathcal{S}$ and issues the ESS power command $a_t$. 
The environment evolves to $s_{t+1}$ and returns a scalar reward that reflects the frequency regulation performance. 
Each time step corresponds to one load disturbance scenario, within which the short-term frequency response is simulated to calculate the reward and the next state. 
The key requirement is that the ESS power command must satisfy the hard constraints at every step:
\begin{equation}
a_t \in \mathcal{A}(t), \quad \forall t\in\mathcal{T},
\label{eq:cmd_feasible}
\end{equation}
Due to modeling uncertainty and the black-box nature of the stability margin evaluation, the transition probability is not available in closed form. 
Therefore, a model free CIS-DRL approach is adopted, where the policy is learned through interactions with a simulator-based environment.

\subsubsection{State space} 

The system state collects the variables relevant to frequency dynamics, ESS operation, and converter-grid interaction stability assessment:
\begin{equation}
\begin{aligned}
s_t = \big\{ &\mathrm{SOC}(t),\, \Delta P(t),\, \Delta f(t),\, v_d(t), \\
            &Q_{\mathrm{ESS}}(t),\, \hat m\big(\boldsymbol{o}(t)\big) \big\}, \quad t \in \mathcal{T},
\end{aligned}
\label{eq:state_def}
\end{equation}
where $\Delta P(t)$ is the power imbalance and $\hat m(\boldsymbol{o}(t))$ is the estimated stability margin at the current operating point. 
These states allow the agent to coordinate active power for frequency regulation and reactive power for voltage support, while accounting for proximity to the converter-grid interaction stability boundary.

\subsubsection{Action space}

In the proposed CIS-DRL environment, the action of the agent is the active and reactive power commands of the ESS:
\begin{equation}
a_t = \{ P_{\mathrm{ESS}}(t),\, Q_{\mathrm{ESS}}(t) \}, \quad t \in \mathcal{T},
\label{eq:action_def}
\end{equation}

The action space bounded by the converter rating and the power limits is given as:
\begin{subequations}\label{eq:action_bounds}
\begin{equation}
-P_{\max}\le P_{\mathrm{ESS}}(t) \le P_{\max},
\end{equation}
\begin{equation}
-Q_{\max} \le Q_{\mathrm{ESS}}(t) \le Q_{\max},
\end{equation}
\end{subequations}
where $Q_{\max}$ is the reactive power capability of the converter.

\subsubsection{Reward function} 

At each time step, the reward function is defined as:
\begin{equation}
r(s_t,a_t) = -\big|\Delta f(t{+}1)\big|.
\label{eq:reward_def}
\end{equation}
where $\Delta f(t{+}1)$ is the frequency deviation at the next step obtained from the discretized swing equation. By maximizing the cumulative discounted reward, the agent is driven to suppress frequency regulation over the entire control horizon.

\subsection{Converter-grid interaction stability constraint}

This subsection develops a CIS-DRL controller by integrating a CIS projection layer into Safe-TD3 to guarantee 100\% interaction stability without violations. 
A double DNN-based stability region evaluation module is first adopted to identify the stable set from PCC measurements, and the projection layer is then used to correct tentative actions into the stability feasible set while obtain optimal operation.

\subsubsection{Double DNN-based stability region evaluation}

\paragraph{$\mathrm{DNN}_{\mathrm{1}}$-Based impedance modeling}
Based on the multi-operating point (MOP) measurement dataset, the dq-axis small signal impedance is first computed and then approximated by a neural network. For each operating point $\boldsymbol{o}$ and perturbation frequency $f_p$, the fundamental PCC voltages and currents corresponding to two linearly independent current injections:
\begin{equation}
\mathbf{V}_{dq}=
\begin{bmatrix}
v_{d1} & v_{d2}\\
v_{q1} & v_{q2}
\end{bmatrix},\qquad
\mathbf{I}_{dq}=
\begin{bmatrix}
i_{d1} & i_{d2}\\
i_{q1} & i_{q2}
\end{bmatrix},
\end{equation}
and the dq-axis impedance is obtained by
\begin{equation}
\mathbf{Z}_{dq}(\boldsymbol{o}, f_p)=
\begin{bmatrix}
Z_{dd} & Z_{dq}\\
Z_{qd} & Z_{qq}
\end{bmatrix}
=\mathbf{V}_{dq}\,\mathbf{I}_{dq}^{-1},
\label{eq:Z_dq}
\end{equation}

The impedance regression network $\mathrm{DNN}_{\mathrm{1}}$ learns the mapping from $(\boldsymbol{o},f_p)$ to ${Z}_{dq}$. The input feature is
\begin{equation}
\boldsymbol{x}=[\,v_d,\ i_d,\ i_q,\ f_p\,]^{\mathrm T},
\end{equation}

To obtain a compact output matrix, the network predicts an 8-dimensional real vector obtained by combining the magnitude and phase of \eqref{eq:Z_dq}:
\begin{equation}
\hat{\boldsymbol z} =
\begin{bmatrix}
\operatorname{vec}\!\big(|\mathbf Z_{dq}|\big) \\[2pt]
\operatorname{vec}\!\big(\angle \mathbf Z_{dq}\big)
\end{bmatrix},
\qquad
\boldsymbol z =
\begin{bmatrix}
\operatorname{vec}\!\big(|\mathbf Z_{dq}|\big) \\[2pt]
\operatorname{vec}\!\big(\angle \mathbf Z_{dq}\big)
\end{bmatrix},
\end{equation}
where $\boldsymbol{z}^{(k)}$ is the target vector formed from \eqref{eq:Z_dq} for the $k$-th sample, and $\hat{\boldsymbol{z}}^{(k)}$ is the corresponding DNN output.

All inputs and outputs are standardized to zero mean and unit variance. $\mathrm{DNN}_{\mathrm{1}}$ is implemented as a fully connected network with two hidden layers and sigmoid activations, using smooth nonlinear activations and a linear output layer. The typical mean squared error (MSE) loss function are calculated by minimizing:
\begin{equation}
\mathcal{L}_{\mathrm{1}}
=\frac{1}{N}\sum_{k=1}^{N}\bigl\|\hat{\boldsymbol{z}}^{(k)}-\boldsymbol{z}^{(k)}\bigr\|_2^{2},
\end{equation}
which uses back-propagation with Adam optimizer. After training, $\mathrm{DNN}_{\mathrm{1}}$ serves as a fast surrogate of $(\boldsymbol{o},f_p)$ over the measured operating domain.

\paragraph{$\mathrm{DNN}_{\mathrm{2}}$-Based stability region evaluation}

With the MOP dq-axis impedance network $\mathrm{DNN}_{\mathrm{1}}$ available, the stability of the system is evaluated in the frequency domain and then learned as a mapping in the operating point space. For a given operating point $\boldsymbol{o}$ and frequency $f$, the converter impedance is obtained from $\mathrm{DNN}_{\mathrm{1}}$ as $\mathbf Z_{dq}(j2\pi f,\boldsymbol{o})$, and the grid impedance at the PCC is represented in the same axis as $\mathbf Z_{g,dq}(j2\pi f)$. The impedance ratio is constructed by:
\begin{equation}
\mathbf L_e(j2\pi f,\boldsymbol{o}) \;=\; \mathbf Z_{g,dq}(j2\pi f)\,\mathbf Z_{dq}^{-1}(j2\pi f,\boldsymbol{o}),
\label{eq:Le}
\end{equation}
where $\mathbf L_e$ plays the role of a multivariable loop function. The eigenlocus for stability assessment is obtained by sweeping the frequency $f$ and tracing the complex eigenvalues $\lambda$ of the impedance ratio:
\begin{equation}
\det\!\big[\lambda\mathbf I-\mathbf L_e(j2\pi f,\boldsymbol{o})\big]=0,
\label{eq:eig_def_clean}
\end{equation}
According to the generalized Nyquist criterion, if $\mathbf L_e(s)$ has $P$ unstable poles, the system is stable if and only if the combined eigenloci encircle the critical point $(-1{+}j0)$ exactly $P$ times over the sweep.

The operating point domain is sampled to obtain the stability dataset:
\begin{equation}
\mathcal D_{\mathrm{stab}}
=\bigl\{\,(\boldsymbol{o}^{(k)},\, y^{(k)})\,\bigr\}_{k=1}^{M},
\qquad
y^{(k)}\in\{0,1\},
\label{eq:D_stab}
\end{equation}
where $y^{(k)}=1$ means system is stable, $y^{(k)}=0$ means system is unstable, which is assigned by the Nyquist criterion using \eqref{eq:Le}--\eqref{eq:eig_def_clean}. The stability network $\mathrm{DNN}_{\mathrm{2}}$ learns the map from operating point to the probability of the stable class, and its parameters are identified by minimizing the cross-entropy loss over $\mathcal D_{\mathrm{stab}}$:
\begin{subequations}\label{eq:CE_loss}
\begin{align}
\mathcal L_{\mathrm{2}}
&= -\frac{1}{M}\sum_{k=1}^{M}\!\left[
y^{(k)}\log \hat p^{(k)} + \big(1-y^{(k)}\big)\log\!\big(1-\hat p^{(k)}\big)
\right], \tag{\theequation a}\label{eq:CE_loss_a}\\[3pt]
\hat p^{(k)} 
&= \mathrm{DNN}_{\mathrm{2}}\!\left(\boldsymbol{o}^{(k)}\right), 
\quad k=1,\ldots,M. \tag{\theequation b}\label{eq:CE_loss_b}
\end{align}
\end{subequations}

With the stability margin available, a safe stability set is defined for a prescribed decision threshold $\tau$ as:
\begin{equation}
\mathcal{S}_{\tau}
= \big\{\, \boldsymbol{o} : \mathrm{DNN}_{\mathrm{2}}(\boldsymbol{o}) \ge \tau \,\big\},
\label{eq:safe_set_tau}
\end{equation}
which collects operating points that $\mathrm{DNN}_{\mathrm{2}}$ classifies as stable
with confidence at least $\tau$. The margin and the associated safe set $\mathcal{S}_{\tau}$ provide a compact description of the stability region in the
$(v_d,i_d,i_q)$ space and will be employed to construct the stability constraint.

\begin{figure*}[t]
\centering
\includegraphics[width=\textwidth]{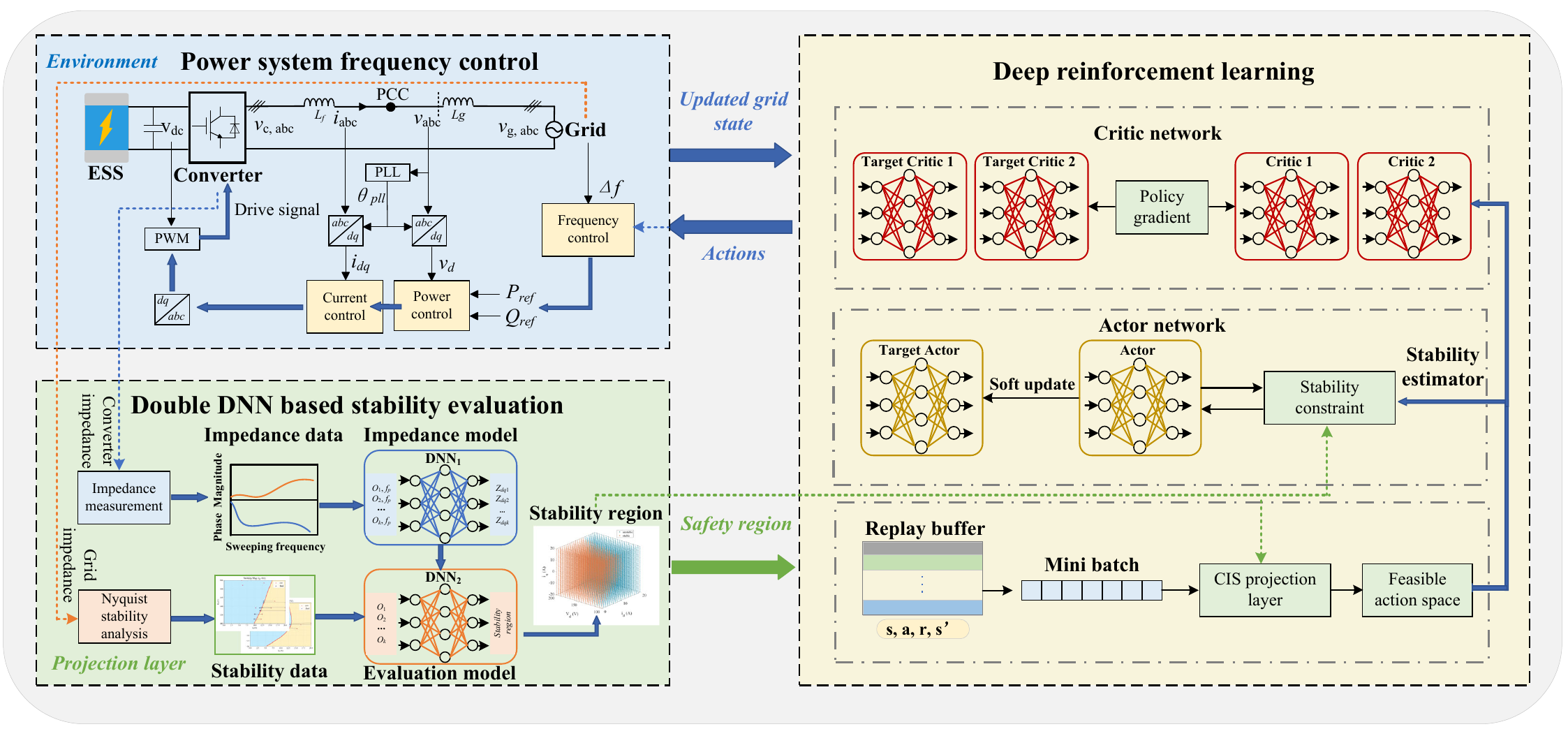}
\caption{Proposed Converter-grid interaction stability guaranteed CIS-DRL method.}
\label{framework1}
\end{figure*}

\subsubsection{CIS projection layer generation }

The projection layer aims to seek the closest stable operating point in $\mathcal{S}_{\tau}$ in
the Euclidean distance. Given an unstable point $\boldsymbol{o}^{\mathrm{u}}_t$, the ideal projection problem is to solve:
\begin{equation}
\boldsymbol{o}^{\mathrm{p}}_t
= \arg\min_{\boldsymbol{o} \in \mathcal{S}_{\tau}}
\big\| \boldsymbol{o} - \boldsymbol{o}^{\mathrm{u}}_t \big\|_2,
\label{eq:projection_problem}
\end{equation}
where $\boldsymbol{o}^{\mathrm{p}}_t$ denotes the projected stable operating point.
Since $\mathcal{S}_{\tau}$ is only available through the black-box classifier
$\mathrm{DNN}_{\mathrm{2}}$, the problem in \eqref{eq:projection_problem} cannot be
solved analytically. In this part, a projection layer that combines outer search radius bisection and a feasibility inner search is designed.

For the outer search, a stable point
$\boldsymbol{o}^{\mathrm{ref}}\in\mathcal{S}_{\tau}$ is maintained, and the initial bounds on the projection radius are chosen as:
\begin{subequations}\label{eq:radius_bounds}
\begin{equation}
r_{\min}^{(0)} = 0,
\end{equation}
\begin{equation}
r_{\max}^{(0)} =
\big\| \boldsymbol{o}^{\mathrm{ref}} - \boldsymbol{o}_t^{\mathrm{u}} \big\|_2,
\end{equation}
\end{subequations}
At iteration $g$, the outer search tests the midpoint radius:
\begin{equation}
r^{(g)} =
\frac{ r_{\min}^{(g)} + r_{\max}^{(g)} }{2},
\label{eq:radius_mid}
\end{equation}
and calls the inner search to check whether there exists at least one stable operating point at distance $r^{(g)}$ from $\boldsymbol{o}_t^{\mathrm{u}}$. If the inner search returns a stable point, the radius is regarded as feasible and the upper bound is updated as $r_{\max}^{(g+1)} = r^{(g)}$. Otherwise, the radius is regarded as infeasible and the lower bound is updated as $r_{\min}^{(g+1)} = r^{(g)}$. This bisection continues until $r_{\max}^{(g)} - r_{\min}^{(g)}$ is smaller than a chosen tolerance.

For the inner search at a fixed radius $r^{(g)}$, a finite set of directions is constructed:
\begin{equation}
\mathcal{D} =
\big\{ \boldsymbol{d}_1,\boldsymbol{d}_2,\ldots,\boldsymbol{d}_M \big\},
\label{eq:direction_set}
\end{equation}
which includes the directions from $\boldsymbol{o}_t^{\mathrm{u}}$ to several stable points and a number of random unit directions in the $(v_d,i_d,i_q)$ space.
For each $\boldsymbol{d}_m\in\mathcal{D}$, a candidate operating point on the sphere is generated as:
\begin{equation}
\boldsymbol{o}_m^{(g)} =
\boldsymbol{o}_t^{\mathrm{u}} +
r^{(g)} \frac{\boldsymbol{d}_m}{\big\|\boldsymbol{d}_m\big\|_2}.
\label{eq:candidate_point}
\end{equation}
and its stability is evaluated by $\mathrm{DNN}_{\mathrm{2}}$. In this way, any tentative action that would lead to an
unstable operating point is automatically redirected inside $\mathcal{S}_{\tau}$, while the stability margin and the admissible set guide the search toward the closest stable region. The detailed procedure of the CIS projection is summarized in Algorithm~\ref{alg:projection}.

\begin{algorithm}
\caption{CIS Projection Layer}
\label{alg:projection}
\begin{algorithmic}[1]
\State \textbf{Input:} initial action $a_t = \{P_{\mathrm{ESS}}(t), Q_{\mathrm{ESS}}(t)\}$, safe stability set $\mathcal{S}_{\tau}$, stable anchor set $\mathcal{A}$, radius tolerance $\varepsilon_r$
\State \textbf{Output:} projected safe action $a_t^{\mathrm{safe}}$
\State Map $(a_t, v_d(t))$ to the operating point $\boldsymbol{o}_t^{\mathrm{u}}$ according to the relations given in \eqref{eq:p_dec}-\eqref{eq:q_dec}
\If{$\mathrm{DNN}_{\mathrm{2}}(\boldsymbol{o}_t^{\mathrm{u}}) \ge \tau$}
    \State Set $a_t^{\mathrm{safe}} = a_t$ and \textbf{return}
\EndIf
\State Choose a reference stable operating point $\boldsymbol{o}^{\mathrm{ref}} \in \mathcal{A}$ and initialise the radius interval with $r_{\min}^{(0)} = 0$ and $r_{\max}^{(0)} = \|\boldsymbol{o}^{\mathrm{ref}} - \boldsymbol{o}_t^{\mathrm{u}}\|_2$, set the current projected point to $\boldsymbol{o}^{\mathrm{ref}}$
\For{$g = 0,1,\ldots,G_{\max}$}
    \If{$r_{\max}^{(g)} - r_{\min}^{(g)} < \varepsilon_r$}
        \State \textbf{break}
    \EndIf
    \State Set the trial radius $r^{(g)}$ as the midpoint of $[r_{\min}^{(g)}, r_{\max}^{(g)}]$
    \State Construct $M$ search directions based on $\mathcal{A}$ and random unit directions in the $(v_d,i_d,i_q)$ space
    \State On the sphere of radius $r^{(g)}$ centred at $\boldsymbol{o}_t^{\mathrm{u}}$, sample candidate operating points along these directions and evaluate their stability
    \If{at least one candidate is classified as stable}
        \State Update $\boldsymbol{o}_t^{\mathrm{p}}$ with a stable candidate and set $r_{\max}^{(g+1)} = r^{(g)}$, $r_{\min}^{(g+1)} = r_{\min}^{(g)}$
    \Else
        \State Set $r_{\min}^{(g+1)} = r^{(g)}$ and keep $r_{\max}^{(g+1)} = r_{\max}^{(g)}$
    \EndIf
\EndFor
\State Interpret the final projected operating point $\boldsymbol{o}_t^{\mathrm{p}}$ and recover the safe action $a_t^{\mathrm{safe}}$
\State \textbf{Return} $a_t^{\mathrm{safe}}$
\end{algorithmic}
\end{algorithm}

\begin{algorithm}
\caption{Converter-grid interaction stability guaranteed Safe-TD3}
\label{alg:modified_td3}
\begin{algorithmic}[1]
\State Initialize critic networks $Q_{\delta_1}$, $Q_{\delta_2}$ and actor network $\pi_{\zeta}$ with random weights
\State Initialize target networks $Q'_{\delta_1}$, $Q'_{\delta_2}$, $\pi'_{\zeta}$ with weights $\delta'_1 \leftarrow \delta_1$, $\delta'_2 \leftarrow \delta_2$, $\zeta' \leftarrow \zeta$
\State Initialize replay buffer $\mathcal{B}$ and reference stable operating point $\boldsymbol{o}^{\mathrm{ref}} \in \mathcal{S}_{\tau}$
\For{episode $=1,\ldots,M$}
    \State Receive initial observation state $s$
    \For{$t=1,\ldots,T$}
        \State Select action $a = \pi_{\zeta}(s) + \epsilon$ and clip it by power and SOC limits
        \State Obtain $v_d(t)$ and map $(a, v_d(t))$ to the tentative operating point $\boldsymbol{o}_t^{\mathrm{u}}$ 
        \If{$\mathrm{DNN}_{\mathrm{2}}(\boldsymbol{o}_t^{\mathrm{u}}) < \tau$}
            \State Apply CIS projection layer (Algorithm~\ref{alg:projection}) with input $a$ and $\boldsymbol{o}^{\mathrm{ref}}$ to obtain projected safe action $a_t^{\mathrm{safe}}$
            \State Update reference stable point $\boldsymbol{o}^{\mathrm{ref}}$ using the projected operating point
        \Else
            \State Set $a_t^{\mathrm{safe}} = a$
        \EndIf
        \State Execute action $a_t^{\mathrm{safe}}$ in the environment and observe reward $r$ and new state $s'$
        \State Store transition $(s, a_t^{\mathrm{safe}}, r, s')$ in $\mathcal{B}$
        \State Sample mini-batch from $\mathcal{B}$ using prioritized replay
        \State Update critic networks by minimizing the TD error
        \State Update actor network by the deterministic policy gradient
        \State Update the target networks with a soft update
    \EndFor
\EndFor
\end{algorithmic}
\end{algorithm}

\subsection{Proposed CIS-DRL method for frequency regulation}

The proposed CIS-DRL method for frequency regulation with guaranteed converter-grid interaction stability is shown in Fig.~\ref{framework1}. 
A simulator-based frequency control environment emulates the grid frequency response and ESS dynamics and returns the system state to the agent. 
The environment consists of a ESS connected to the grid with a synchronous $dq$-axis control structure, including a PLL for synchronization and PI inner loops for current control. 
Frequency regulation is achieved by modulating the ESS active power output.

The converter-interfaced system is treated as a black-box at the operation level. The grid impedance is obtained from PCC measurements via impedance scanning, and interaction stability is assessed in the frequency domain using the Nyquist criterion. 
Based on stability assessment results over multiple operating points, the double DNN-based stability evaluation module is trained to characterize the stability region.

The learned stability region is used as an online feasibility query for the DRL controller. At each decision step, the actor first generates a tentative action with exploration noise:
\begin{equation}
a_t^{\mathrm{raw}} = \pi_{\zeta}(\hat{s}_t) + \varepsilon_t,
\end{equation}
where $\varepsilon_t$ is Gaussian noise, and the result is clipped to the action bounds determined by the converter capability.
The noise is then corrected by the CIS projection layer to produce an admissible command before execution. 
In this way, stability feasibility is enforced at every step by construction, and the executed action satisfies both stability feasibility and operational limits.

Based on this interaction mechanism, the underlying actor--critic learning follows the TD3 structure with two critics and one actor. 
The target networks of the critics and the actor are softly updated as:
\begin{subequations}\label{eq:td3_soft_update}
\begin{equation}
\delta'_j = \kappa \delta_j + (1-\kappa)\,\delta'_j, \qquad j=1,2,
\end{equation}
\begin{equation}
\zeta' = \kappa \zeta + (1-\kappa)\,\zeta',
\end{equation}
\end{subequations}
where $\kappa \ll 1$ is the soft update rate, which is further reduced in the later stage
of training to obtain more stable targets. 
The environment returns $(s_t,r_t,s_{t+1})$ after applying the admissible action, and the transition is stored in the replay buffer for off-policy updates. 
Algorithm~\ref{alg:modified_td3} summarizes the resulting CIS-DRL procedure with the interaction stability projection embedded at each time step.

\section{Case Study and Validations}
\begin{figure}[htbp]
\centerline{\includegraphics[width=0.48\textwidth]{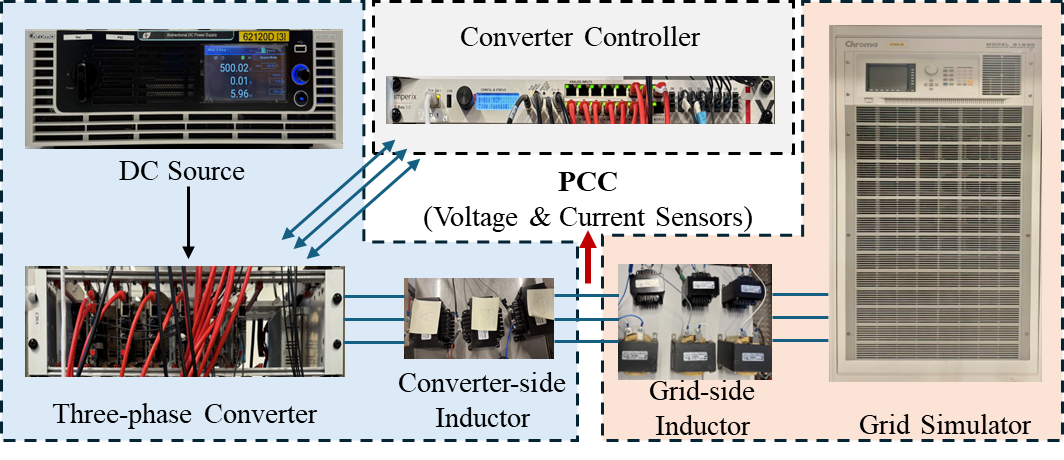}}
\caption{The experiment setup of the converter-grid interaction system.}
\label{exp}
\end{figure}

To validate the proposed Converter-grid interaction stability guaranteed CIS-DRL method, case studies are carried out on a voltage source converter (VSC)-based ESS integrated grid, which is displayed in Fig.~\ref{exp}. The DC source supplies the three-phase VSC, whose AC side is connected to the grid through a converter-side inductor. The PCC is located between the converter-side and grid-side inductors, where voltage and current are measured for control and analysis purposes. For the grid side, the system is connected to the grid simulator via a grid-side inductor, enabling a flexible emulation of different grid conditions.
The detailed parameters of the tested grid connected VSC are listed in Table~\ref{tab:vsc_param}.

The validation is organized by linking stability analysis, online feasibility enforcement, and closed-loop frequency regulation performance. First, the double DNN-based Converter-grid interaction stability region is visualized and examined, and the projection behavior is demonstrated to guarantee stability feasibility at each control step. Next, the operation level frequency regulation performance of the proposed method is compared with baseline case, no ESS integration, and the case that with ESS integration but no stability consideration. The performance includes frequency deviations, ESS charging and discharging trajectories, SOC dynamic, and converter current profiles. Finally, the hardware experiments are conducted to demonstrates real world effectiveness.

Python 3.8 is used to implement the DRL environment and the agent, and the neural networks are trained with the PyTorch 2.1.1 framework on a desktop computer equipped with a 1.80 GHz 13th Gen Intel Core i7-1365U processor and 32 GB RAM, running 64-bit Windows 10 Pro. For the hardware experiment, an imperix controller is used to control the converter, and a 62120D programmable DC power source is employed to supply the DC voltage for the converter, while the Chroma programmable power source 61845 is used as the grid simulator.

\begin{table}
\begin{center}
\caption{VSC Parameters}
\label{tab:vsc_param}
\begin{tabular}{| c | c | c |}
\hline
Variable & Description & Value \\
\hline
$V_{dc}$ & DC-side voltage & 500 V \\
$f_{\mathrm{sw}}$ & Switching frequency & 10 kHz \\
$f_s$ & Sampling frequency & 10 kHz \\
$f_0$ & Nominal grid frequency & 50 Hz \\
$L_f$ & Converter-side inductor & 3 mH \\
$L_g$ & Grid-side inductor & 20 mH \\
$v_{\mathrm{base}}$ & AC-side base voltage & 110 V \\
$K_{pi}$ & Proportional gain of current controller & 13.09 \\
$K_{ii}$ & Integral gain of current controller & 3426 \\
$K_{p\mathrm{pll}}$ & Proportional gain of PLL & 5.63 \\
$K_{i\mathrm{pll}}$ & Integral gain of PLL & 2591 \\
\hline
\end{tabular}
\end{center}
\end{table}

\begin{figure}[htbp]
\centerline{\includegraphics[width=0.43\textwidth]{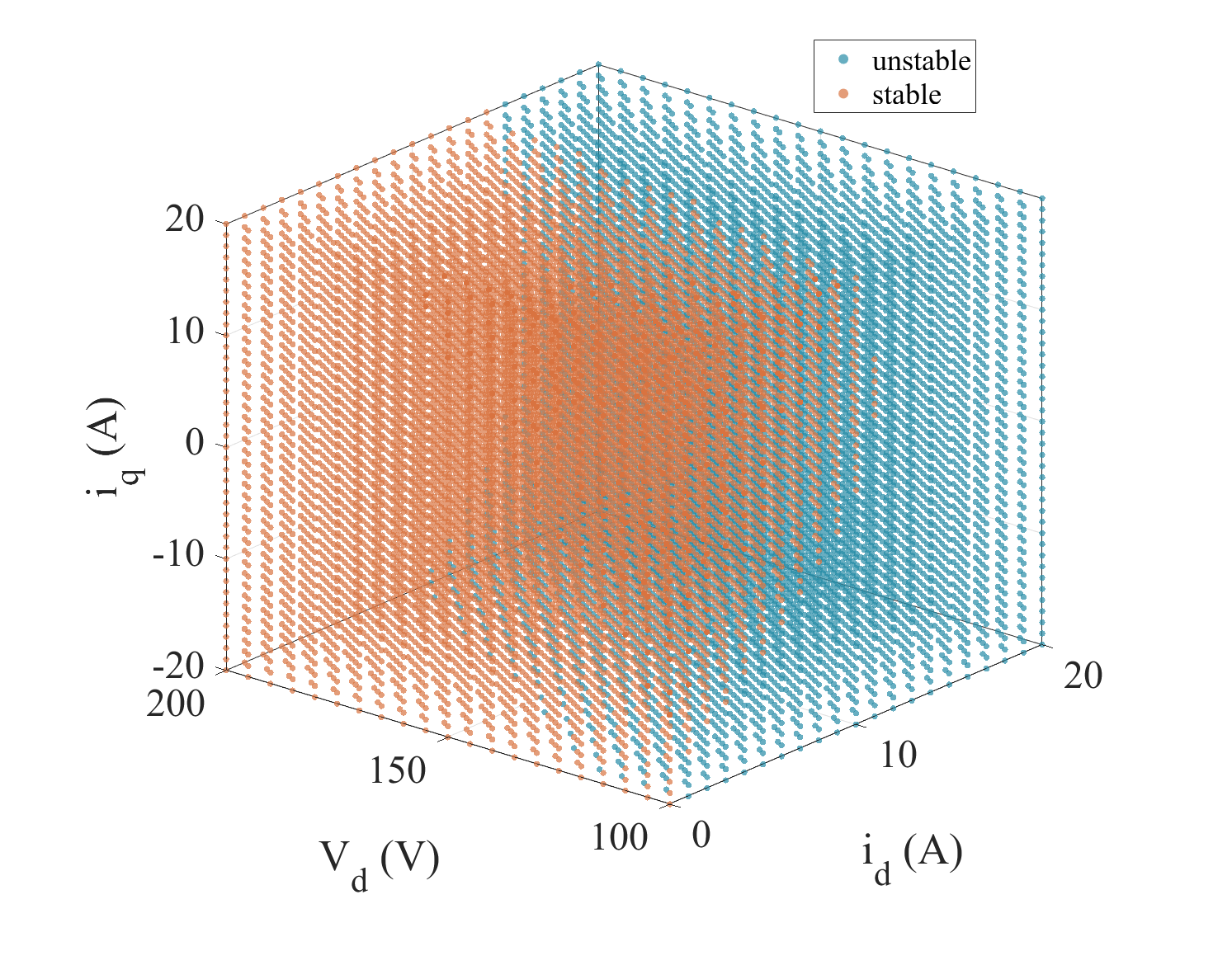}}
\caption{Calculated converter-grid interaction stability region.}
\label{stabilityregion}
\end{figure}

\begin{figure}[htbp]
\centerline{\includegraphics[width=0.43\textwidth]{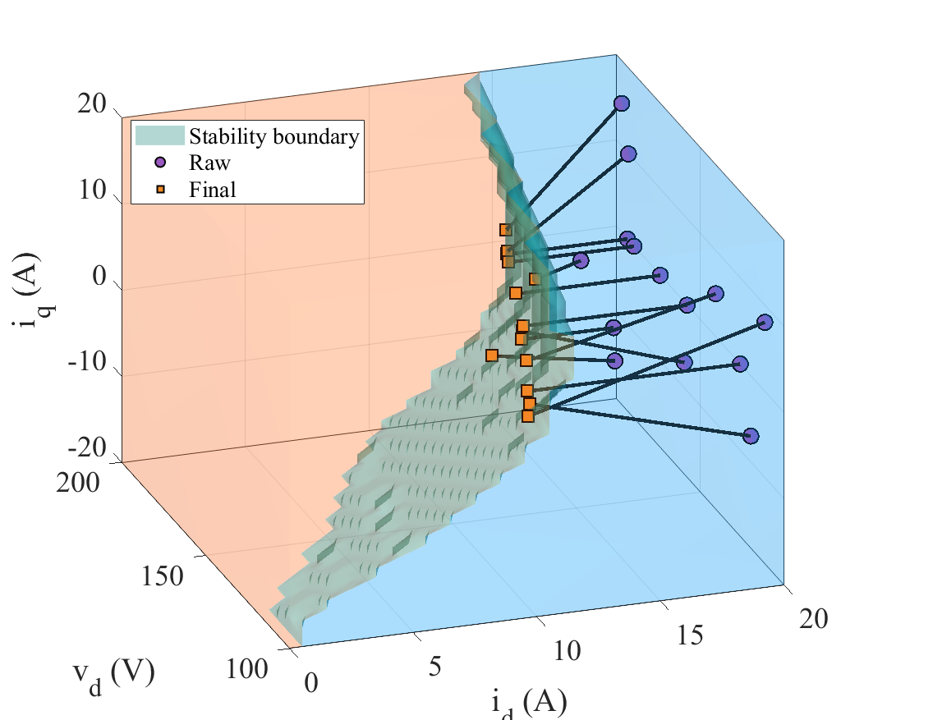}}
\caption{Converter-grid interaction stability with projection results.}
\label{stabilityregion_project}
\end{figure}

\subsection{Converter-grid interaction stability region evaluation and projection results}

\begin{figure}[htbp]
\centerline{\includegraphics[width=0.53\textwidth]{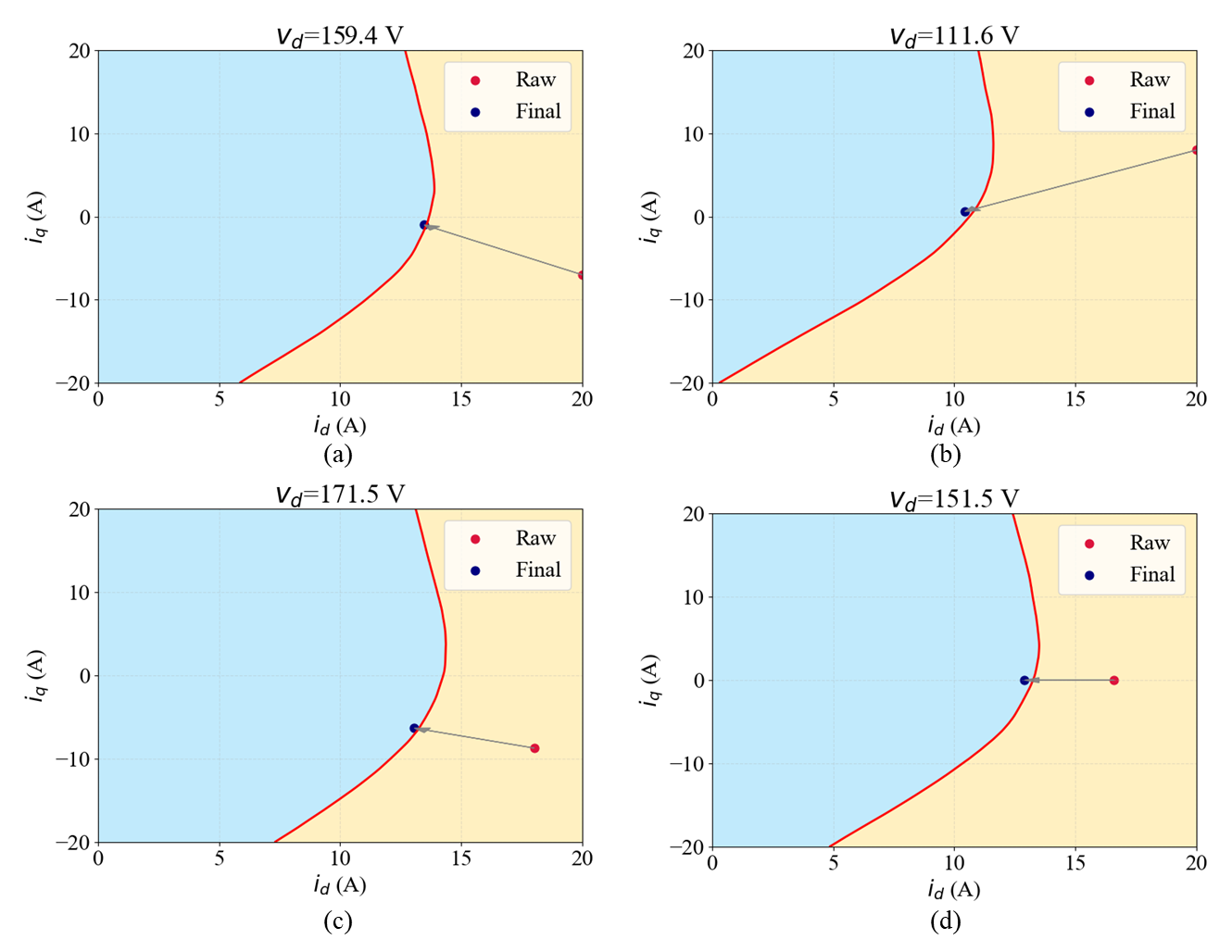}}
\caption{The operation points in different $v_d$: (a) $v_d$=159.4V, (b) $v_d$=111.6V, (c) $v_d$=171.5V, (d) $v_d$=151.5V.}
\label{stabilityall}
\end{figure}

The learned converter-grid interaction stability region in the $(v_d,i_d,i_q)$ space is shown in Fig.~\ref{stabilityregion}, where blue dots denote unstable region and orange dots denote stable region. 
Based on the learned stability region, the projection behavior is demonstrated in Fig.~\ref{stabilityregion_project}, which provides a three dimensional view of the stability region together with the projection results. Purple dots indicate the original operating points, and the orange dots indicate the projected operating points. It can be observed that inadmissible tentative points are redirected into the stability region before execution, consistent with the projection design. 
These results confirm that the double DNN model provides reliable stability region evaluation, which will be employed in the Safe DRL to support the frequency regulation during closed-loop operation.

To better visulize the projection result, Fig.~\ref{stabilityall} presents four slices at the typical voltage levels corresponding to selected decision steps ($v_d = 159.4$~V, $111.6$~V, $171.5$~V, and $151.5$~V). In each subplot, it can be observed that the projection always moves the point along the shortest direction towards the boundary in the $(i_d,i_q)$ plane, consistent with the design of the projection layer.

\begin{figure}[htbp]
\centerline{\includegraphics[width=0.5\textwidth]{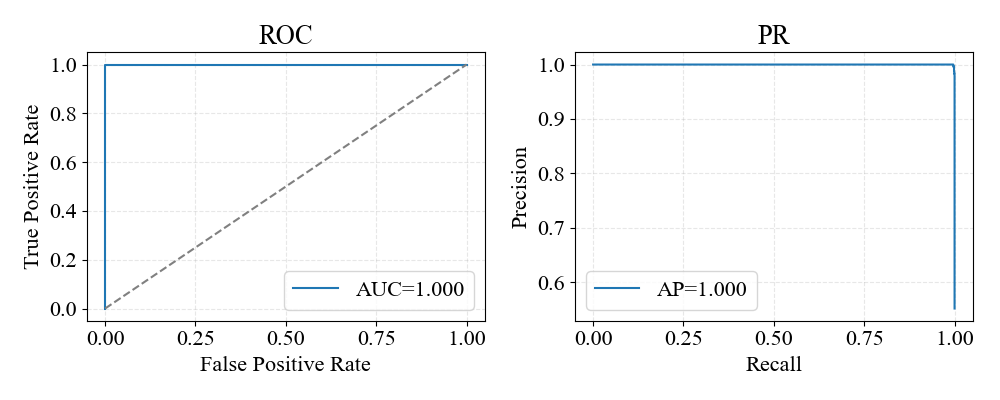}}
\caption{$\mathrm{DNN}_{\mathrm{2}}$ training profile: receiver operating characteristic (ROC) and precision recall (PR) curves.}
\label{DNNtraining}
\end{figure}

The details of DNN neural networks training is then introduced. 
The training data are generated by frequency scanning the VSC model introduced in Section II. The scanning ranges are chosen to cover the typical operating domain of the test system. In particular, the fundamental frequency perturbation varies from $1$~Hz to $100$~Hz with a step of $1$~Hz. The $d$-axis current $i_d$ is swept from $0$~A to $20$~A in steps of $1$~A, the $q$-axis current $i_q$ ranges from $-20$~A to $20$~A with a step of $10$~A, and the $d$-axis voltage $v_d$ is varied between $100$~V and $200$~V in steps of $10$~V. This procedure results in $18,081$ labelled samples, which are split into training and validation sets and are used to train both $\mathrm{DNN}_{\mathrm{1}}$ and $\mathrm{DNN}_{\mathrm{2}}$.
For $\mathrm{DNN}_{\mathrm{2}}$, the ROC and PR curves on the validation set are shown in Fig.~\ref{DNNtraining}, which shows that both stable and unstable classes are correctly identified with high precision and recall. The area under the ROC curve (AUC) and the average precision (AP) are used to quantify the classification performance.
The final validation accuracy reaches $99.64\%$. Both networks are trained offline from operating data and are then used online to provide fast stability evaluation for the safe DRL training.

\begin{figure}[htbp]
\centerline{\includegraphics[width=0.5\textwidth]{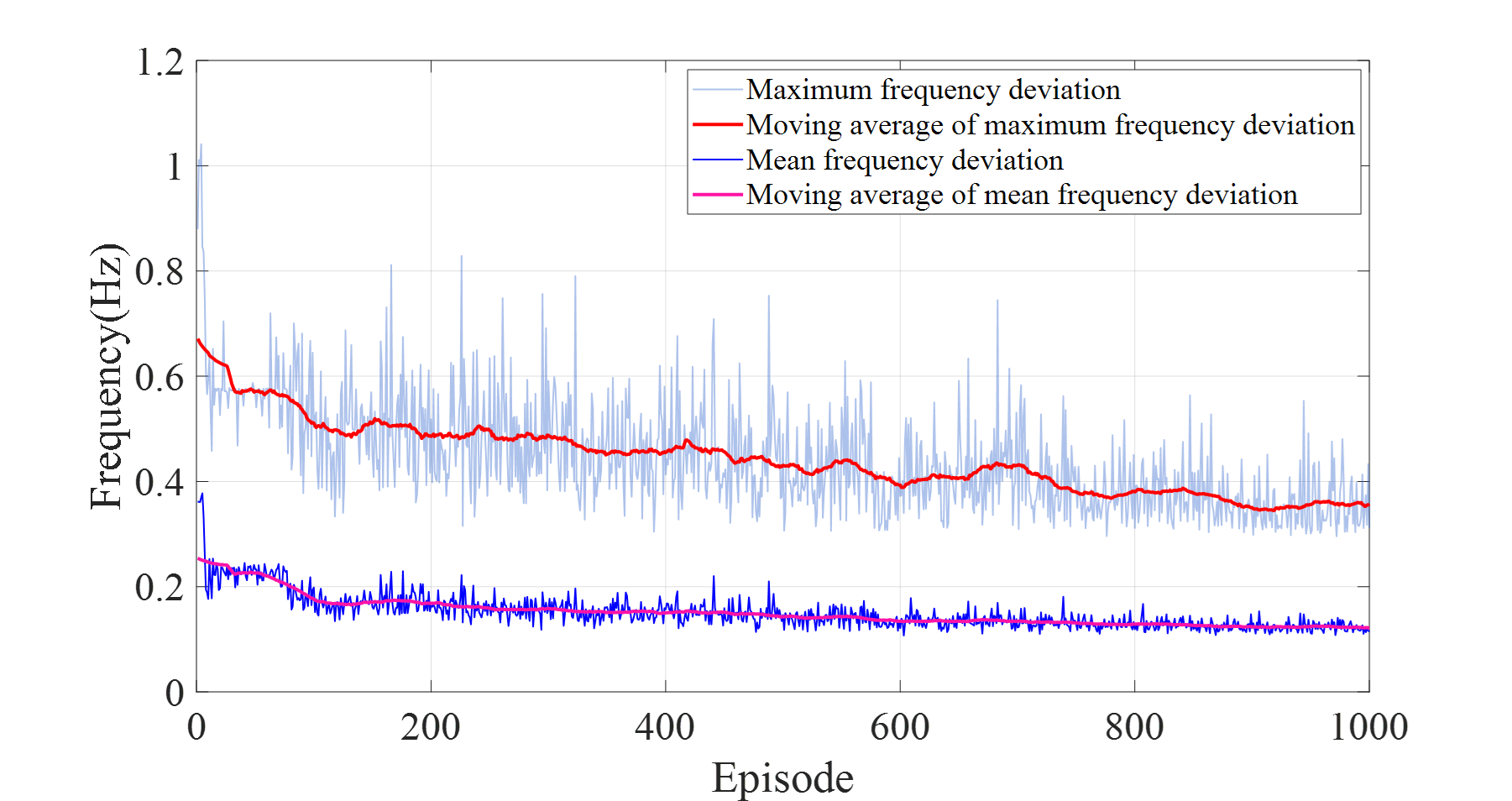}}
\caption{Maximum and mean frequency deviation during training process.}
\label{frequencydeviation}
\end{figure}

\subsection{Converter-grid interaction stability guaranteed frequency regulation}

In the CIS-STD3 algorithm, one actor and two critic networks are adopted in a standard twin delayed structure, and the CIS projection layer is applied to every tentative action before it is sent to the environment. The actor is a three layer fully connected network with 400 and 300 hidden neurons, layer normalization and ReLU activation, and a $\tanh$ output scaled by the action bounds to generate the ESS active power command. Two critics are trained in parallel to reduce overestimation and have the same architecture. The hyperparameters for the algorithm are listed in Table~\ref{tab1}.

\begin{figure}[htbp]
\centerline{\includegraphics[width=0.5\textwidth]{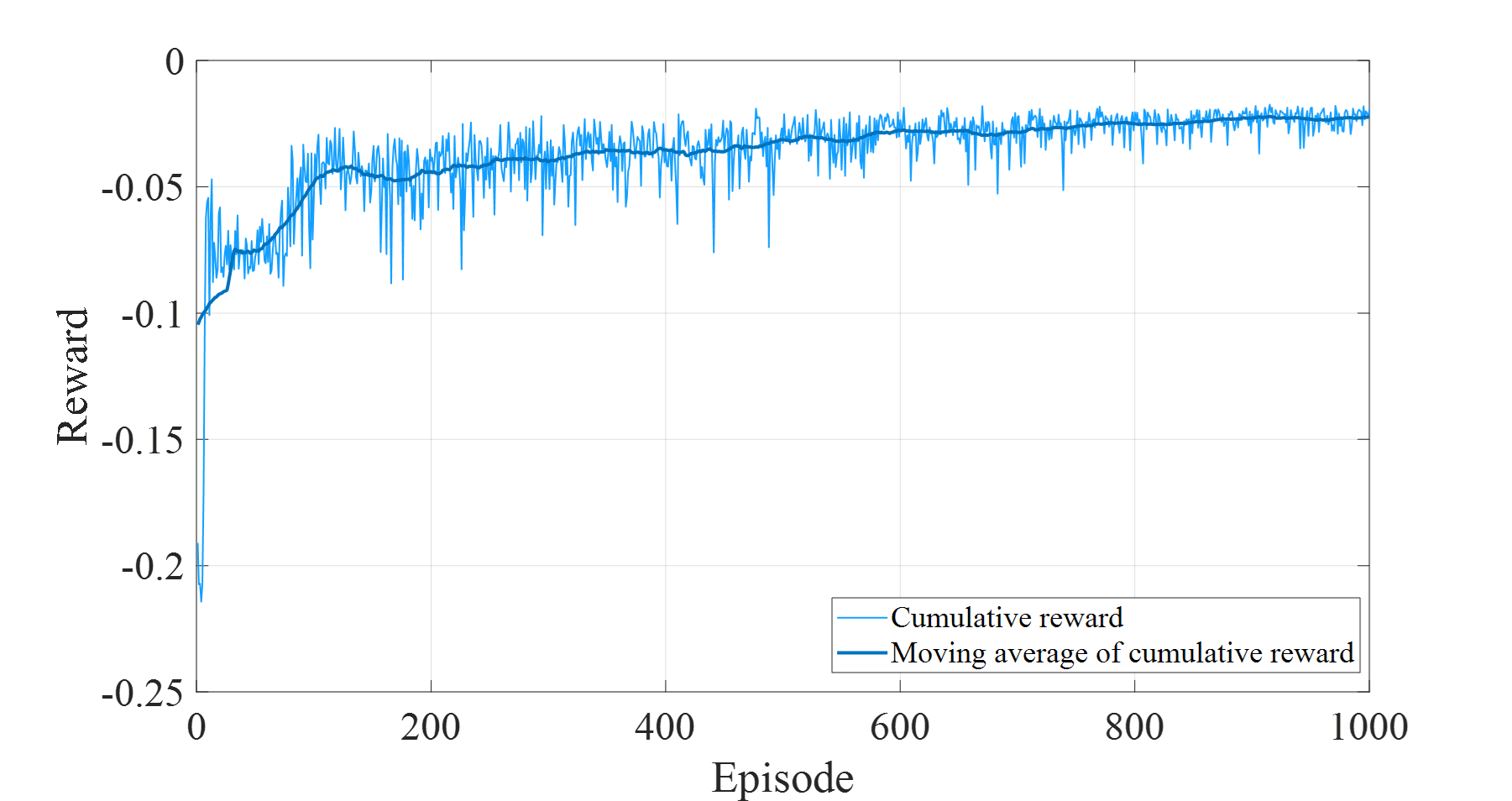}}
\caption{Cumulative reward of the DRL agent during the training process.}
\label{reward}
\end{figure}

\begin{table}
\begin{center}
\caption{Hyperparameters of the Training Process.}
\label{tab1}
\begin{tabular}{| c | c |}
\hline
Hyperparameters&Value\\
\hline
Batch size& 256\\
\hline
Replay buffer& 1e6\\
\hline
Initial exploration rate&0.9\\ 
\hline
Initial actor-network learning rate&1e-4\\
\hline
Initial critic-network learning rate& 1e-4\\
\hline
Target policy smoothing coefficient& 0.1\\
\hline
Discount factor&0.99\\
\hline 
\end{tabular}
\end{center}
\end{table}

The learning behaviour of the CIS Safe-TD3 agent is illustrated in
Figs.~\ref{frequencydeviation} and \ref{reward}. Fig.~\ref{frequencydeviation} shows the maximum frequency deviation and the mean frequency deviation over the dynamic response, together with their moving averages. Both indices exhibit a clear downward trend, indicating the learned policy progressively improves the damping of frequency excursions caused by load disturbances.
Fig.~\ref{reward} shows the cumulative reward per episode and the moving average. Starting from strongly negative values in the early exploration stage, the cumulative reward increases steadily and gradually converges to a plateau close to zero.

\begin{figure}[htbp]
\centerline{\includegraphics[width=0.5\textwidth]{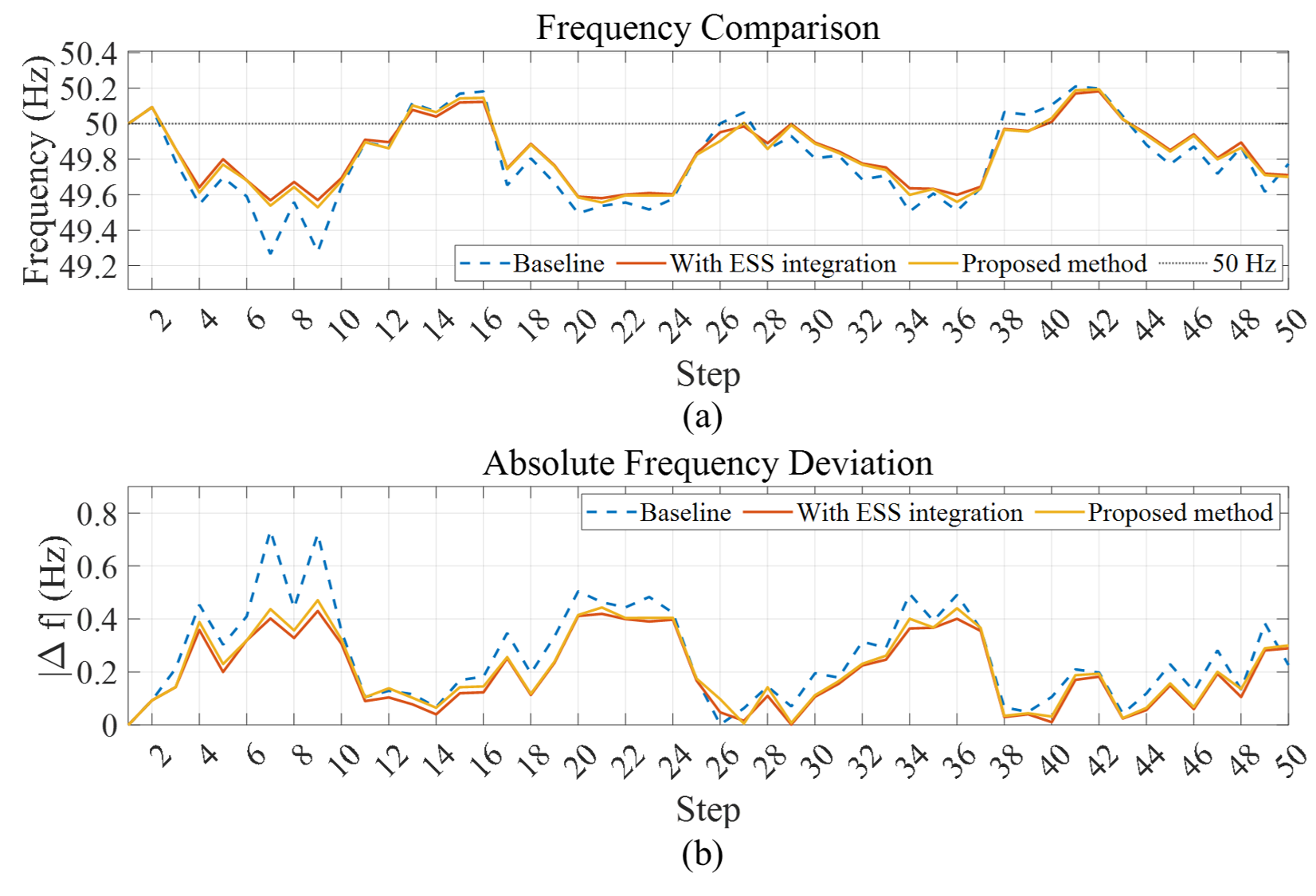}}
\caption{Frequency comparison between base case without ESS integration, Case 1 with ESS integration but no stability projection, and Case 2 with stability projection: (a) frequency comparison, (b) absolute frequency deviation.}
\label{frequencycomparisonall}
\end{figure}

\begin{figure}[htbp]
\centerline{\includegraphics[width=0.45\textwidth]{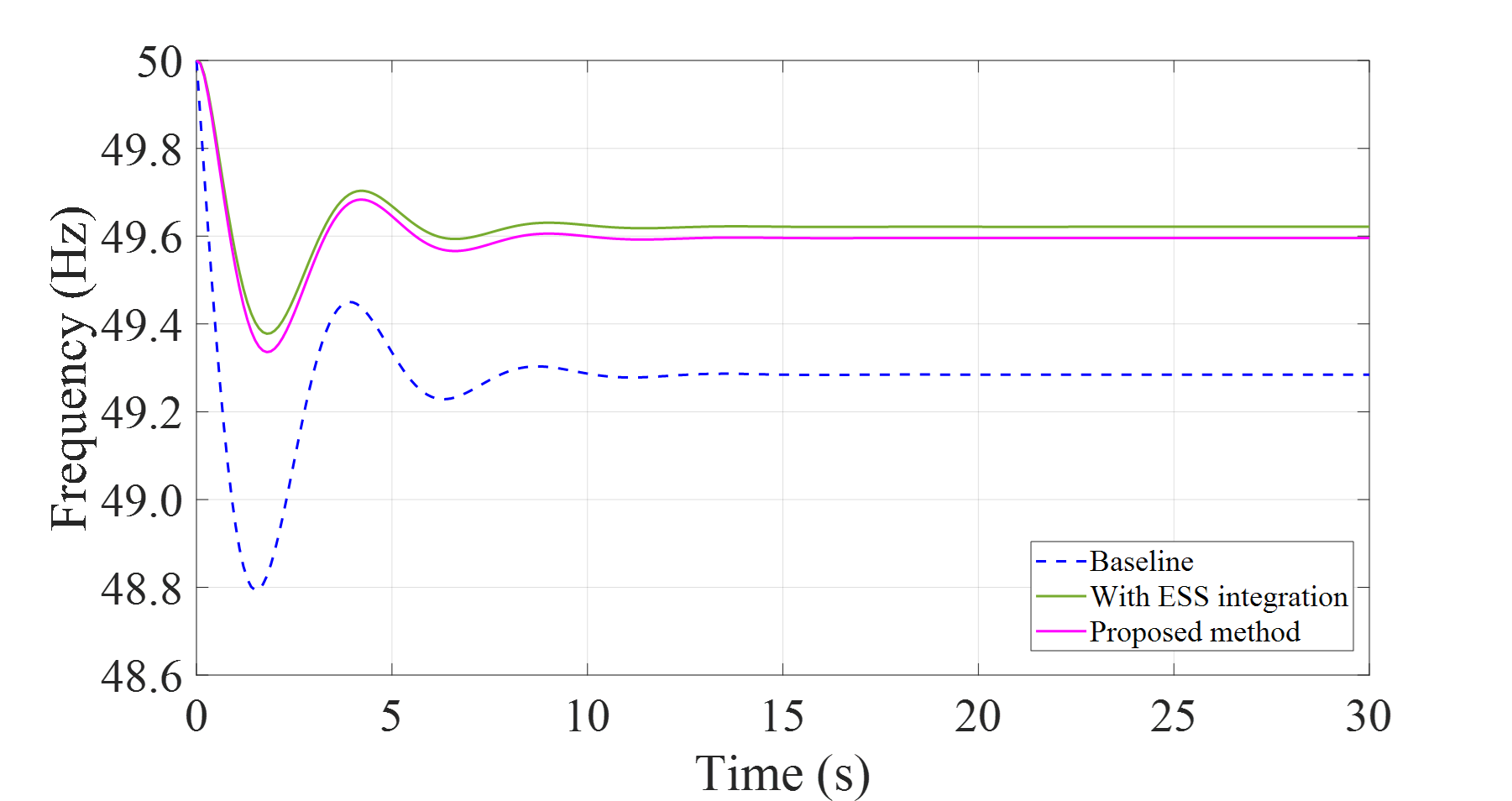}}
\caption{Dynamic frequency responses of Step 7 in Fig.~\ref{frequencycomparisonall} for the three cases.}
\label{dynamicfrequency}
\end{figure}

Fig.~\ref{frequencycomparisonall} compares the frequency trajectories over 50 steps for three cases. Each step denotes one decision interval in the DRL based frequency regulation process. The baseline case does not consider ESS integration. The second case shows the conventional method, which has ESS integration but without converter-grid interaction stability guaranteed. The third is the proposed method with ESS and CIS projection. As shown in Fig.~\ref{frequencycomparisonall}(a), without ESS support the system frequency exhibits large excursions from the nominal 50~Hz, while both ESS based cases noticeably improve the dynamic response. The corresponding absolute deviations in Fig.~\ref{frequencycomparisonall}(b) shows frequency deviation. The maximum frequency deviation in the base case reaches $0.7338$~Hz, whereas it is reduced to $0.4017$~Hz when the ESS is controlled without projection and to $0.4371$~Hz with the proposed CIS-DRL method. Fig.~\ref{dynamicfrequency} shows the dynamic frequency response of example step 7 in Fig.~\ref{frequencycomparisonall}. 
Although at some control steps the case without projection yields the smallest maximum deviation, it ignores the converter-grid interaction stability and will drive the operating point into the unstable region, which makes such a policy impractical for real deployments. In contrast, the proposed method slightly sacrifices frequency performance but guarantees that every ESS action lies inside the learned stability region.

\begin{figure}[htbp]
\centerline{\includegraphics[width=0.5\textwidth]{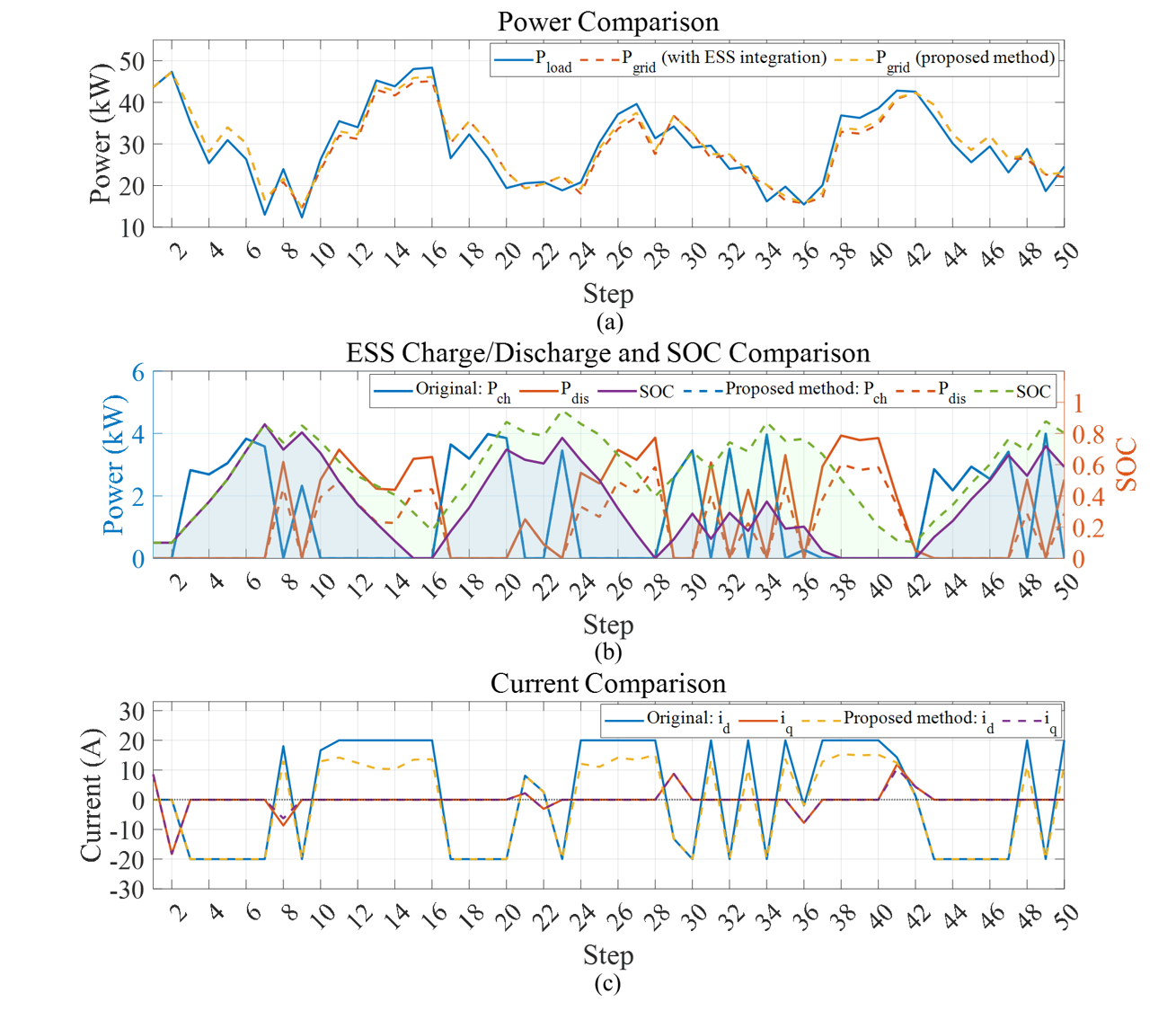}}
\caption{Power demand, charging/discharging power, and current comparison between base case without ESS integration, Case 1 with ESS integration but no stability projection, and Case 2 with ESS integration and stability projection: (a) power comparison, (b) ESS power output and SOC comparison, (c) current comparison.}
\label{powersoccomparisonall}
\end{figure}

To demonstrate that our proposed method guarantee the operation limits, Fig.~\ref{powersoccomparisonall}(a) compares the active power at the load side and at the grid side for the three cases. Fig.~\ref{powersoccomparisonall}(b) shows the corresponding ESS charge, discharge powers and SOC evolution. With stability projection, the charging and discharging power is moderated, leading to a smoother SOC trajectory and avoiding aggressive swings that would push the operating point towards the edge of the stability region.
The impact on converter currents is highlighted in
Fig.~\ref{powersoccomparisonall} (c). Without projection, the $d$ and
$q$-axis currents exhibit large spikes and approach the rated current boundary at several steps, reflecting the unconstrained search. When the CIS projection layer is activated, both $i_d$ and $i_q$ are kept within a tighter envelope and follow much smoother profiles, while still providing effective power support. This indicates that the proposed method not only mitigates frequency fluctuations, but also enforces practical current limits and converter–grid interaction stability during the entire control process. In Figs. \ref{frequencycomparisonall}-\ref{powersoccomparisonall}, each step denotes one decision interval in the simulated frequency regulation process.

\begin{figure*}[htbp]
\centerline{\includegraphics[width=0.99\textwidth]{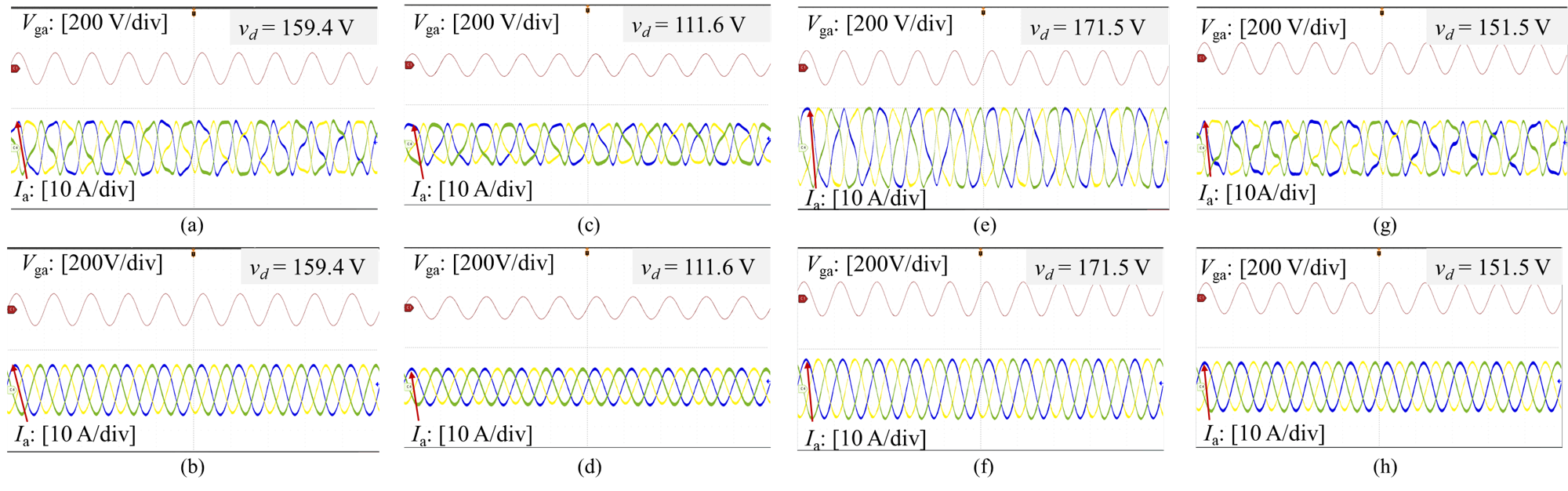}}
\caption{Experimental results of the VSC-based ESS integrated grid: (a) conventional method without stability projection when $v_d = 159.4$~V, (b) proposed method when $v_d = 159.4$~V, (c) conventional method when $v_d = 111.6$~V, (d) proposed method when $v_d = 111.6$~V, (e) conventional method when $v_d = 171.5$~V, (f) proposed method when $v_d = 171.5$~V, (g) conventional method when $v_d = 151.5$~V, (h) proposed method when $v_d = 151.5$~V.}
\label{exp1_stabilityall}
\end{figure*}

Finally, to demonstrate the effectiveness of the proposed method to guarantee converter-grid interaction stability, the experimental results of measured grid-phase voltage $V_{ga}$ and converter output currents $I_a$, $I_b$, and $I_c$ are shown in Fig.~\ref{exp1_stabilityall}. These points correspond to the raw and final points in Fig.~\ref{stabilityall}.
Taking Fig.~\ref{exp1_stabilityall}(e) and (f) as examples, the operating point $(i_d,i_q) = (18.0~\mathrm{A}, -8.7~\mathrm{A})$ in (e) is unstable, while (f) is the projected stable point with $(i_d,i_q) = (13.1~\mathrm{A}, -6.3~\mathrm{A})$.
For the unstable operating points, the three-phase currents exhibit oscillations and waveform distortion, indicating loss of small signal stability of the converter–grid interaction. In contrast, after the CIS projection of the proposed method, for operating point~b the currents are nearly balanced and following the uniform distribution, confirming the stable operation.

\section{Conclusion}

This paper proposes a safe DRL based frequency regulation method that guarantees converter-grid interaction stability for ESS integrated systems under incomplete converter information and for wide operating point variations. Using a learned stability region as a safe action set, the CIS-DRL agent applies an interaction stability projection mechanism to correct unsafe commands before execution, ensuring stepwise stability while minimizing frequency deviation. Case studies show that the proposed method improves frequency regulation and prevents unstable operating points by mapping infeasible actions into stable region, with ESS current and power trajectories consistently remaining within feasible limits. Experimental results indicate a strong potential for practical deployment in real-world converter-dominated grids under uncertainty and varying operating conditions.

\end{document}